\documentclass[twocolumn,aps,prd,superscriptaddress,nofootinbib,preprintnumbers,10pt]{revtex4-1}
\usepackage{graphicx}
\usepackage{amsmath}
\usepackage{bm}
\usepackage{slashed}
\usepackage{epsfig}
\usepackage{amsfonts}
\usepackage{epstopdf}
\usepackage{hyperref}
\usepackage[subfigure]{graphfig}
\usepackage{braket}
\usepackage{tikz}
\usepackage{hhline}
\usepackage[normalem]{ulem} 
\usepackage{soul}

\newcommand{\ben}{\begin{eqnarray}}
\newcommand{\een}{\end{eqnarray}}

\newcommand{\bef}{\begin{figure}[!htp]}
\newcommand{\eef}{\end{figure}}

\newcommand{\nn}{\nonumber}

\newcommand{\half}{{\textstyle{\frac{1}{2}}}}

\def\be{\begin{equation}}
\def\ee{\end{equation}}
\newcommand{\bea}{\begin{eqnarray}}
\newcommand{\eea}{\end{eqnarray}}
\newcommand{\lb}{\label}
\newcommand{\req}[1]{(\ref{#1})}

\def\ba{\begin{linenomath*}\begin{equation}}
\def\ea{\end{equation}\end{linenomath*}}

\definecolor{forestgreen}{rgb}{0.13, 0.55, 0.13}

\usepackage{soul}

\hypersetup{colorlinks, linkcolor = [rgb]{0, 0, 0.5}, citecolor = [rgb]{0,0.0,1.0}, urlcolor = [rgb]{0,0.0,0.5}}

\begin{document}

\preprint{  JLAB-THY-21-3454}
\preprint{ SLAC-PUB-17612}

\title{ Gluon matter distribution in the proton and pion from extended holographic light-front QCD}

\newcommand*{\COSTARICA}{Laboratorio de F\'isica Te\'orica y Computacional, Universidad de Costa Rica, 11501 San Jos\'e, Costa Rica}\affiliation{\COSTARICA}
\newcommand*{\HEIDELBERG}{Institut f\"ur Theoretische Physik der Universit\"at, D-69120 Heidelberg, Germany}\affiliation{\HEIDELBERG}
\newcommand*{\SDU}{Key Laboratory of Particle Physics and Particle Irradiation (MOE), Institute of Frontier and Interdisciplinary Science, Shandong University, Qingdao, Shandong 266237, China}\affiliation{\SDU}
\newcommand*{\WM}{ Department of Physics, William \& Mary, Williamsburg, Virginia 23187, USA}\affiliation{\WM}
\newcommand*{\JLAB}{Thomas Jefferson National Accelerator Facility, Newport News, VA 23606, USA}\affiliation{\JLAB}
\newcommand*{\SLAC}{SLAC National Accelerator Laboratory, Stanford University, Stanford, CA 94309, USA}\affiliation{\SLAC}

\author{Guy~F.~de~T\'eramond}\affiliation{\COSTARICA}
\author{H.~G.~Dosch}\affiliation{\HEIDELBERG}
\author{Tianbo~Liu}\email{liutb@sdu.edu.cn}\affiliation{\SDU}
\author{Raza~Sabbir~Sufian}\email{sufian@jlab.org}\affiliation{\WM}\affiliation{\JLAB}
\author{Stanley~J.~Brodsky}\affiliation{\SLAC}
\author{Alexandre~Deur}\affiliation{\JLAB}

\collaboration{HLFHS Collaboration}

%\date{\today}

\begin{abstract}

The holographic light-front QCD framework provides a unified nonperturbative description of the hadron mass spectrum, form factors and quark distributions. In this article we extend holographic QCD in order to describe the gluonic distribution in both the proton and pion from the coupling of the metric fluctuations induced by the spin-two Pomeron with the energy momentum tensor in anti--de Sitter space, together with constraints imposed by the Veneziano model{\color{blue},} without additional free parameters. The gluonic and quark distributions are shown to have significantly different effective QCD scales.
 
\end{abstract}

\maketitle

%%%%%%%%%%%%%%%%%%%%%%%%

\section{Introduction \lb{intro}}

The gluonic composition of hadrons plays a key role in understanding the confining phase of quantum chromodynamics (QCD), which is still an unresolved issue in modern particle physics.  A key nonperturbative feature of color-confining hadron dynamics is the intrinsic gluon distribution which exists in hadrons over a time scale independent of the resolution of the external probe.  The coupling of the rank-two energy-momentum tensor (EMT), the tensor which couples to gravity~\cite{Kobzarev:1962wt, Weinberg:1964ew, Pagels:1966zza},  provides  fundamental constraints on the quark and gluon generalized parton distribution functions (GPDs) of a hadron~\cite{Mueller:1998fv, Ji:1996ek, Radyushkin:1997ki}.
Gravitational form factors (GFFs), the hadronic matrix elements of the EMT, describe the coupling of a hadron to the graviton and thus provide information on the dynamics of quarks and gluons within hadrons due to the internal shear forces and pressure distributions of the quarks and gluons~\cite{Polyakov:2002yz, Polyakov:2018zvc, Burkert:2018bqq}. In this article, we present an extended holographic light-front (LF) QCD framework for studying the gluon GFFs and provide predictions for the intrinsic gluon distributions of hadrons without introducing additional parameters.

In addition to the role of gluons as fundamental constituents and as the glue binding quarks into hadrons, the knowledge of gluon distributions within hadrons is also essential for the understanding of Higgs boson production~\cite{Georgi:1977gs} and particle production cross sections at small-momentum fraction $x$. The near threshold production of heavy vector quarkonium~\cite{Brodsky:2000zc}, such as the $J/\psi$ and $\Upsilon$, is dominated by multiple gluon interaction identified with Pomeron exchange~\cite{Brodsky:1989jd}: It is expected to shed light on the QCD  trace  anomaly  at the origin of the proton mass~\cite{Kharzeev:1998bz}. On the other hand, the intrinsic gluon parton distribution function (PDF) of the pion,  the lightest QCD bound state, is of particular theoretical interest for understanding nonperturbative aspects of QCD,  such as the connection of the spontaneous breaking of chiral symmetry and confinement~\cite{DeTeramond:2021jnn}.  Because of the special role of the pion in QCD, there have been sustained efforts and proposals to explore its gluon distribution~\cite{Aguilar:2019teb}, along with that of the nucleon, one of the main goals of the upcoming Electron Ion Collider~\cite{Accardi:2012qut, AbdulKhalek:2021gbh}.

Holographic light-front  QCD (HLFQCD), a nonperturbative framework based on the gauge/gravity correspondence~\cite{Maldacena:1997re}  and its light-front  holographic mapping~\cite{Brodsky:2006uqa,deTeramond:2008ht,Brodsky:2014yha}, has the remarkable feature that it reproduces, within its expected precision,
the hadronic spectra with the minimal number of parameters: the confining scale $\lambda$ and effective quark masses.  The effective confining interaction for mesons and baryons is determined by an underlying superconformal algebraic structure~\cite{Fubini:1984hf, deTeramond:2014asa, Dosch:2015nwa}, which leads to unexpected connections across the full hadron spectrum~\cite{Brodsky:2020ajy}. It reproduces the structure of hadronic spectra as predicted by dual models, most prominently the Veneziano model~\cite{Veneziano:1968yb} with its typical  features;  linear Regge trajectories with  a universal slope and the existence of ``daughter trajectories''.

The form factors (FFs) obtained within the HLFQCD framework can be expressed by Euler-Beta functions~\cite{deTeramond:2018ecg,Zou:2018eam}, a feature also predicted by the generalized Veneziano model~\cite{Ademollo:1969wd,Landshoff:1970ce}, which includes the electromagnetic (EM) current and the  FFs; 
$F_{\rm EM}(t) \sim B\left(\gamma, 1 - \alpha_\rho(t)\right)$, where $\alpha_\rho(t)$ is  the trajectory of the  $\rho$-vector-meson,  coupling to the quark current in the hadron. We emphasize that, in HLFQCD,  the parameter $\gamma$, related to the falloff of the EM FF at large momentum transfer $t$, is not arbitrary, but fixed by the twist of the Fock state, $\gamma = \tau -1$, consistent with the exclusive counting rules~\cite{Brodsky:1973kr,Matveev:1973ra}. The twist $\tau$ is the number of constituents $N$,  $\tau = N$ ($\tau = N + L$ for LF orbital angular momentum $L$).

The form of the quark distributions in the hadrons is heavily constrained by the analytic properties of the FFs. Furthermore, very natural assumptions, such as the incorporation of the inclusive-exclusive relation at large longitudinal momentum fraction  $x$~\cite{Drell:1969km}, allowed us to predict, in a very satisfactory way, the quark distributions in mesons and  nucleons~\cite{deTeramond:2018ecg,Liu:2019vsn}, as well as the strange-antistrange  and the charm-anticharm asymmetries~\cite{Sufian:2018cpj,Sufian:2020coz} in the nucleon. Notably, the  preliminary NNPDF4.1 global analysis~\cite{Rojo:2021gdq} indicates a valencelike intrinsic charm contribution $\ket{uudc\bar{c}}$ in the nucleon, consistent with the intrinsic charm-anticharm asymmetry computed in~\cite{Sufian:2020coz}. In addition to higher twist, the sea quark distributions have distinctive Regge trajectories from the EM current couplings to the sea: $\alpha_{\phi}(t)$ and $\alpha_{J/\Psi}(t)$, respectively for the strange and charm sea in the proton~\cite{Sufian:2018cpj,Sufian:2020coz}. Likewise, the $\ket{uudg}$ and $\ket{u\bar{d}g}$ Fock states should provide the leading contributions to the intrinsic gluon distributions of, respectively, the proton and pion coupled to a spin-two current with its characteristic Reggeized $t$-channel particle exchange.

 In this article we extend our previous framework by incorporating gluonic matter with significantly different quarkonic and gluonic scales. Several models for nonperturbative QCD, see e.g.~\cite{DiGiacomo:2000irz},  predict that the static potential between quarks, $V_{\bar{q}q}$, or between gluons, $V_{gg}$, are substantially different, and scale  with $N_C$ like the quadratic Casimir operator between the fundamental and adjoint representations, $\frac{V_{gg}}{V_{\bar{q}q}} \sim \frac{2 N_C^2}{N_C^2 -1}$, a scaling confirmed by lattice calculation~\cite{Bali:2000un}. It is in qualitative accordance with the smaller slope of the Pomeron as compared to the Reggeons, $\alpha_P \ll \alpha_\rho$, therefore with different scales $\lambda_q$ and $\lambda_g$  for quarkonic and gluonic matter with $\lambda_g \gg \lambda_q$.

The rest of this paper is organized as follows. In Sec. \ref{PEXGFF} we study the perturbation of the anti--de Sitter (AdS) metric by an external spin-two current which couples to the EMT in  the bulk. It allows us to identify the tensor Pomeron in the asymptotic physical boundary space with the induced metric fluctuations in AdS, and thus to determine the scale of the gluonic matter from the infrared deformation of the AdS metrics in terms of the Pomeron slope. By further imposing the structure of the generalized Veneziano amplitudes for a spin-two current, we are able to incorporate the Pomeron intercept in the analytic expression for the gravitational form factor $A^g(Q^2)$. Our results for the GFF for the proton and the pion are compared with recent lattice QCD computations. In Sec. \ref{GDF} we extend the relation between FFs  and GPDs~\cite{Mueller:1998fv, Ji:1996ek, Radyushkin:1997ki} found in~\cite{deTeramond:2018ecg} to compute the gluon distribution functions. Our DGLAP evolved predictions for the proton and pion are then compared with global fits. Some final comments are included in Sec. \ref{CaO}.

%%%%%%%%%%%%%%%%%%%%%%%%%%%%%%%

%%%%%%%%%%%%%%%%%%%%%%%%%%%%%%%

\section{Pomeron exchange and gravitational form factors \label{PEXGFF}}

As pointed out in~\cite{Zyla:2020zbs}, soft interactions  play an important role in high-energy collisions.  Especially,  the trajectory  for diffractive processes, the Pomeron, has a dominant role at small-angle high-energy scattering, which is beyond the applicability of perturbative QCD. Since the early days of QCD,  Pomeron exchange was associated with two (or more) gluons~\cite{Gribov:1984tu, Badelek:1992gs, Donnachie:2002en}.  The  Pomeron couples as a rank-two tensor to hadrons~\cite{Adamczyk:2012kn, Ewerz:2013kda, Ewerz:2016onn, Lebiedowicz:2016zka,  Britzger:2019lvc} and couples strongly to gluons. It remains unclear  whether  there exists a relation between the soft~\cite{Donnachie:1992ny} and hard~\cite{Kuraev:1977fs, Balitsky:1978ic, Kirschner:1983di} Pomerons and their crossover regime. It may be sufficient to consider only the soft Pomeron if one looks into the intrinsic gluon component of the  nucleon structure functions, except, perhaps, for extremely small $x$~\cite{Kuraev:1977fs, Balitsky:1978ic, Kirschner:1983di, Mueller:1985wy, Kovchegov:1999yj}.
Therefore we use  the soft Pomeron of Donnachie and Landshoff~\cite{Donnachie:1992ny}  with the effective Regge trajectory,
\begin{align} \label{RP}
\alpha_{P}(t) = \alpha_{ P}(0) + \alpha'_{P} t,
\end{align}
which is interpreted in QCD as a $J^{PC} = 2^{++}$ bound state of two gluons with  intercept $\alpha_P(0) \simeq 1.08$  and  slope $\alpha’_P \simeq  0.25 \, {\rm GeV}^{-2}$~\cite{Zyla:2020zbs}.  Pomeron exchange is identified as the graviton of the dual  AdS theory~\cite{Brower:2006ea, Cornalba:2008sp, Domokos:2009hm, Brower:2010wf, Costa:2012fw, Costa:2013uia, Amorim:2021gat} and the first hadronic state on the Pomeron trajectory is identified with the $2^{++}$ glueball~\cite{glueballs}.

We start with the AdS gravity action with five-dimensional coordinates $x^M = \left(x^\mu, z \right)$
\begin{align}  \label{SG}
S_G[g] &= -  \frac{1}{4}
\int \! d^5x  \sqrt{g} \, e^{\varphi_g(z)} {\left( \mathcal{R} - \Lambda\right)},
\end{align}
modified by an exponential dilaton term $e^{\varphi_g(z)}$ in the string frame. The metric determinant is $g$, $\mathcal{R}$ is the scalar curvature and $\Lambda = - 6/R^2$, with $R$ the AdS radius.
We consider the perturbation of the action \req{SG} by an arbitrary external source at the AdS asymptotic boundary which propagates inside AdS space and couples to the EMT~\cite{Abidin:2008ku, Brodsky:2008pf}. By performing a  small  deformation of the AdS  metric about its AdS background, $g_{M N}  \to g_{M N} + h_{MN}$, we obtain the effective action $S_{\rm eff}[h, \Phi] = S_g[h] + S_{i}[h, \Phi]$
\begin{align}  \label{Sg}
S_{g}[h] &= -  \frac{1}{4}
\int \! d^5x  \sqrt{g} \, e^{\varphi_g(z)} \big(\partial_L h^{M N} \partial^L h_{M N}  \nn \\ 
& \hspace{60pt} - \frac{1}{2} \partial_L h \, \partial^L h \big),\\  \lb{Si}
S_{i}[h, \Phi] &=  \frac{1}{2} \int \! d^5x  \sqrt{g} \, h_{M N} T^{M N}(\Phi),
\end{align}
where  $\Phi$ represents the matter field content.  In deriving \req{Sg}, we use the harmonic gauge $\partial_L h_M^L = \half \partial_Mh$, $h \equiv h_L^L$. The interaction term \req{Si}  describes the coupling of the 
EMT with the graviton probe in AdS--which we identify here with the Pomeron.

We consider the propagation of $h_{M N}$ with components along Minkowski coordinates and impose the additional gauge condition $\partial_M h = 0$. From \req{Sg} we obtain the linearized Einstein  equations
\begin{align} \label{hWE}
- \frac{z^3}{e^{\varphi_g(z)} }\partial_z \Big( \frac{e^{\varphi_g(z)}}{z^3} \partial_z h_\mu^{\, \nu} \Big)
+ \partial_\rho  \partial^\rho h_\mu^{\, \nu} = 0,
\end{align}
where $h_{\mu \nu}$ couples to the transverse and traceless part of the EMT  in \req{Si}. The boundary limit of the graviton probe is a plane wave solution with polarization indices along the physical coordinates 
$h_\mu^{\, \nu}(x, z \to 0) = \epsilon_\mu^{\, \nu} e^{- i q \cdot x}$,
$q^2 = - Q^2 < 0$. Thus, $h_\mu^{\, \nu}(x, z ) = \epsilon_\mu^{\, \nu} \, e^{- i q \cdot x} H(q^2, z)$, 
with $H(q^2= 0, z) = H(q^2, z = 0) = 1$. 

As an example, we represent the quark matter content of a pion by a scalar field $\Phi$  with  action
\begin{align} \lb{Sq}
S_q[\Phi] =  \int \! d^5 x  \sqrt{g} e^{\varphi_q(z)} \! \left( g^{M N} \partial_M \Phi^*\partial_N \Phi -  \mu^2 \Phi^* \Phi \right),
\end{align}
modified by the dilaton term $e^{\varphi_q(z)}$. It represents a pion with physical mass $P_\mu P^\mu  = M^2_\pi$ and AdS mass $(\mu R)^2  = - 4$~\cite{Brodsky:2014yha}. The EMT  follows from \req{Sq},  with 
\begin{align}
T_{M N}  = \partial_M \Phi^* \partial_N \Phi + \partial_N \Phi^* \partial_M \Phi.
\end{align}
It leads from the interaction term \req{Si} to the transition amplitude
\begin{align} \lb{hT}
 \int \! d^5 x \, \sqrt{g}\, h_{M N}  \left(
\partial^{M} \Phi_{P'}^* \partial^{N} \Phi_P+
\partial^{N} \Phi_{P'}^* \partial^{M} \Phi_P \right).
\end{align} 
 
The action $S_q[\Phi]$ \req{Sq} is modified by  the dilaton term  $e^{\varphi_q (z)}= e^{\lambda_q z^2}$ with $\lambda_q = 1/4 \alpha_\rho’  \simeq (0.5 \, {\rm GeV})^2$, specific to the light front mapping to physical $(3+1)$-dimensional space in  HLFQCD~\cite{Brodsky:2014yha} and the constraints imposed by the superconformal algebraic structure~\cite{Brodsky:2020ajy}.   On the other hand, the spin-two action $S_g[h]$ \req{Sg} describing the propagation of a gravitational fluctuation is only constrained by gauge invariance, therefore it is modified by a negative soft-wall dilaton profile~\cite{Karch:2006pv}  $e^{\varphi_g (z)} = e^{- \lambda_g z^2}$ with $\lambda_g = 1/4  \alpha_P' \simeq 1 \, {\rm GeV}^2$ from the Pomeron slope. The interaction term $S_i[h, \Phi]$ \req{Si} has no deformation term and does not introduce an additional scale. For a soft-wall profile $\varphi_g(z) = - \lambda_g z^2$ the solution to \req{hWE} is given by
\begin{align} \lb{H}
H(a, \xi) & = \Gamma \left(2 + a\right) U\left(a, -1, \xi \right)  \\
& =  a \left(2 + a\right) \
 \int_0^1 dx  \, x^{a- 1} (1-x)  e^{- \xi x(1-x)}  \nn ,
\end{align}
where $a = {Q^2}/{4 \lambda_g}$, $\xi = \lambda_g z^2$, and $U(a,b,z)$ is the Tricomi confluent hypergeometric function.

The usual expression of the GFF from the hadronic matrix elements of the EMT
\begin{align}
\left\langle P' \left\vert T_\mu^{\, \nu} \right\vert P \right\rangle 
=  \left( P^\nu P'_\mu + P_\mu  P'^\nu \right) A(Q^2),
\end{align}
follows from extracting the delta function from momentum conservation  in \req{hT}. We obtain for $A(Q^2)$~\cite{Abidin:2008ku, Brodsky:2008pf}
\begin{align}  \lb{A}
A_\tau(Q^2)  =    \int_0^\infty \frac{dz}{z^3} \, H(Q^2, z) \Phi_\tau^2(z),
\end{align}
with $\Phi_\tau(z)$,  
the twist-$\tau$ hadron bound-state solution~\cite{Brodsky:2014yha}  and normalization $A_\tau(0) = 1$. 

To compute the gluon GFF, $A^g_\tau(Q^2)$, the Pomeron is assumed to couple mainly to the constituent gluon~\cite{Adamczyk:2012kn, Ewerz:2013kda, Ewerz:2016onn, Lebiedowicz:2016zka,  Britzger:2019lvc}. The lowest twist is the $\tau = 4$ Fock state $|uudg\rangle$ in the proton and the twist $\tau = 3$ state $|u \bar d g \rangle$ in the pion, both containing an intrinsic gluon.  The effective physical scale for this process is the scale of the Pomeron which couples to the constituent gluon over a distance $\sim 1 / \sqrt{\alpha’_P}$ , therefore described by the wave function $\Phi^g_\tau(z)  \sim z^\tau e^{ - \lambda_g z^2/2}$~\cite{Brodsky:2014yha}.  Upon substituting in~\req{A} the $x$-integral representation of the bulk to boundary propagator \req{H} we find

\begin{align} \lb{GFFA}
A^g_\tau(Q^2) = \frac{1}{N_\tau} B \left(\tau - 1, 2 - \alpha_P(Q^2)\right),
\end{align}
 with $N_\tau =  B \left(\tau - 1, 2 - \alpha_P(0)\right)$, the result  of the generalized Veneziano model~\cite{Veneziano:1968yb, Ademollo:1969wd, Landshoff:1970ce} for a spin-two current.

In writing \req{GFFA} we have assumed that only the dilaton profile describing Pomeron exchange sets the scale when computing the gluon GFFs and GPDs, while only the dilaton corresponding to Reggeon exchange needs to be considered when finding the EM FFs and quark GPDs~\cite{deTeramond:2018ecg}. In \req{GFFA} we have also shifted the Pomeron intercept to its physical value $\alpha_P(0) \approx 1$, since the expression which follows from the holographic result \req{A} leads to a zero intercept. This procedure, which has been successfully applied to our treatment of the EM FF~\cite{deTeramond:2018ecg}, maintains the analytic properties of the GFF: It amounts to a displacement of the timelike poles 
of the exchanged particles in the $t$-channel~\cite{deTeramond:2018ecg, Afonin:2021cwo}. In fact, for integer twist the GFF \req{GFFA} is expressed as a product of $\tau -1$ timelike poles located at 
\begin{align} \lb{M2RE}
   - Q^2 = M^2_n = \frac{1}{\alpha_{P}'}\left(n + 2 - \alpha_{P}(0)\right), 
\end{align}
the radial excitation spectrum of the spin-two Pomeron.
The lowest state in this trajectory, the $2^{++}$,  has the mass $M \simeq 1.92$ GeV, compared with the lattice results  $M \simeq (2.15 - 2.4)$ GeV~\cite{Morningstar:1999rf, Meyer:2004jc, Meyer:2004gx, Zyla:2020zbs}.  

 %%%%%%%%%%%%%%%%%%%%%
 \begin{figure}[htp]
	\centering
	\setlength\belowcaptionskip{-8pt}
 	\includegraphics[width=1.00\columnwidth]{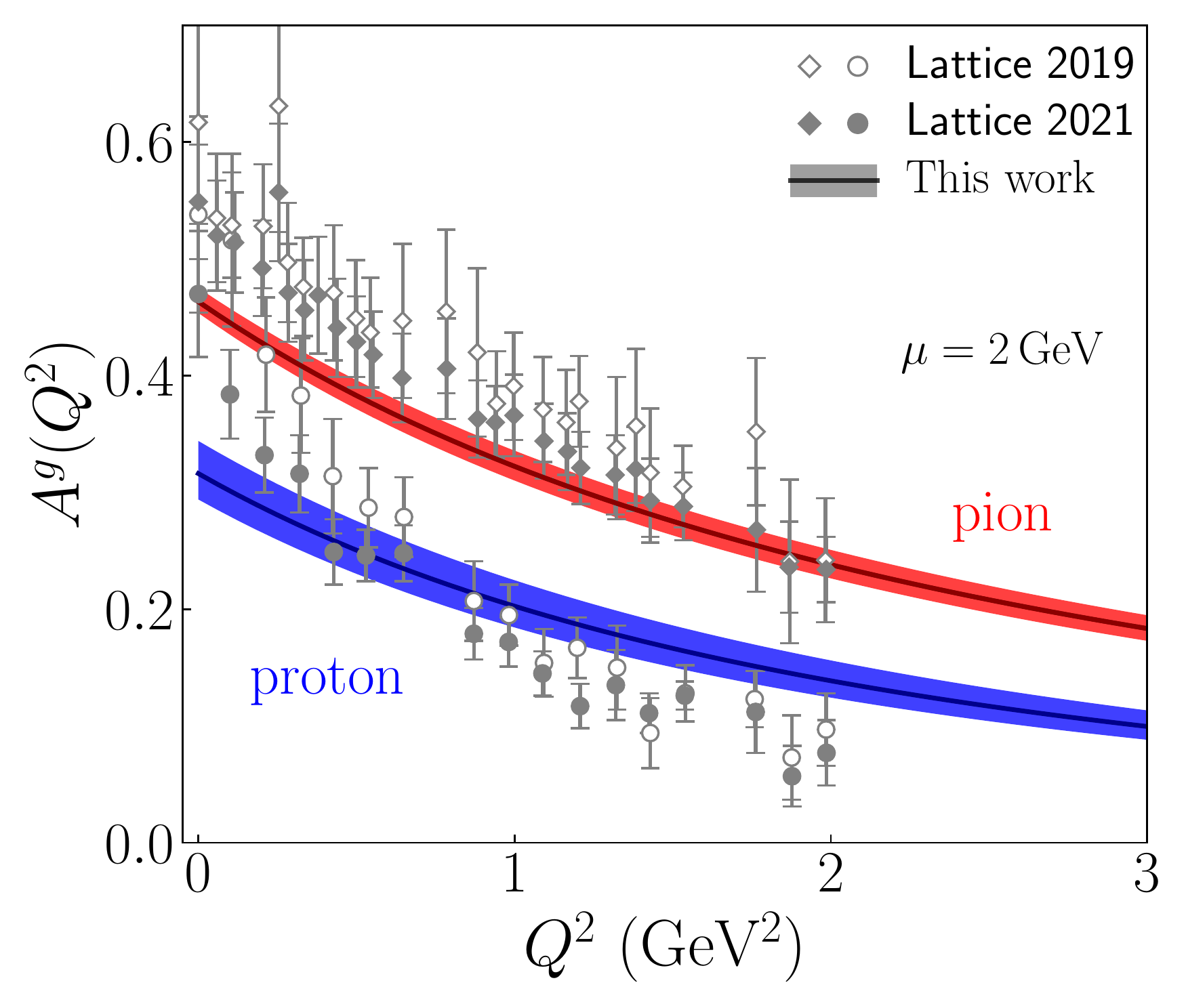}
 	\caption{Gluon  gravitational form factor $ A^g(Q^2)$ of the proton (blue) and the pion (red) in comparison with lattice QCD calculations in Refs.~\cite{Shanahan:2018pib, Pefkou:2021fni}. The value $ A^g(0)$ corresponds to the momentum fraction carried by gluons:  $0.316\pm0.025$ for the proton and $0.463\pm0.011$ for the pion at $\mu=2\,\rm GeV$. The bands indicate the uncertainty from the variation of $\lambda_g$ by $\pm 5\%$ and the normalization from the momentum sum rule.} \label{fig:GFF}
 \end{figure}
%%%%%%%%%%%%%%%%%%%%%

The predictions for the GFF, $ A^g(Q^2)$, for the nucleon and pion are presented in Fig.~\ref{fig:GFF}. We find for the gluon mass  squared radius 
\begin{align}
\langle r_g^2 \rangle = \frac{6}{A^g(0)} \frac{ d A^g(t)}{dt}\Big\vert_{t = 0},
\end{align} 
$\langle r_{g}^2 \rangle_p  =  2.93 / \lambda_g =(0.34 \, {\rm fm})^2$ and $\langle r_{g}^2 \rangle_\pi = 2.41 / \lambda_g = (0.31 \, {\rm fm})^2$ for the proton and pion,  indicating a gluon-mass distribution concentrated in a rather small region  compared with the spread of the charge~\cite{Zyla:2020zbs}, and also smaller than the proton mass radius found in ~\cite{Abidin:2008ku, Guo:2021ibg, Kharzeev:2021qkd, Mamo:2021krl}. The normalization used in Fig.~\ref{fig:GFF} is discussed below.

%%%%%%%%%%%%%%%%%%%%%%%%%%%%%%%%%%%%%%%

%%%%%%%%%%%%%%%%%%%%%%%%%%%%%%%%%%%%%%%

\section{Gluon distribution functions \lb{GDF}}

Recent calculations of the gluon distribution functions in the  hadrons have been performed using holographic approaches, such as in~\cite{Watanabe:2019zny, Mamo:2019mka, Lyubovitskij:2020xqj, Lan:2021wok}. Following the unified approach advanced in~\cite{deTeramond:2018ecg, Liu:2019vsn} we determine in the present work  the unpolarized  gluon distributions in the nucleon and pion without introducing any new free parameters.  In the nonperturbative domain, low virtuality gluons interact strongly with each other to generate the color confinement potential and one cannot distinguish individual gluon quanta. At higher virtualities constituent gluons appear as new degrees of freedom.  The lowest gluonic Fock state  of the proton is $ \ket{uudg}$ and, for simplicity, we consider this Fock state to be the dominant contribution to the intrinsic gluon distribution. 

Using \req{GFFA} and the integral representation of the Beta function, the gluon GFF $A_\tau(t)$ can be written in the reparametrization invariant form
\begin{align}  \lb{Atauw}
A^g_\tau(t) = \frac{1}{N_\tau}  \int_0^1 dx \, w'(x) w(x)^{1 - \alpha_{_P}(t)} \big[1- w(x)\big]^{\tau-2},
\end{align}
provided that $w(x)$ satisfies the constraints $w(0) = 0$,  $w(1) = 1$  and $w'(x) \ge 0$~\cite{deTeramond:2018ecg, Liu:2019vsn}. 
The GFF can also be expressed as the first moment  of the gluon GPD at zero skewness, $H^g_\tau(x, t) \equiv H^g_\tau(x ,\xi=0,t)$,
\bea \label{AHg}
A^g_\tau(t) &=& \int_0^1 x \, dx  H^g_\tau(x,t)  \\
&=&\int_0^1 x \, dx \, g_\tau(x) e^{t f(x)}, 
\eea
where $f(x)$ is the profile function and $g_\tau(x)$ is the collinear gluon PDF of twist-$\tau$. Comparing \eqref{AHg} with the holographic expression \eqref{Atauw} we find that both functions, $f(x)$ and $g_\tau(x)$, are determined in terms of the reparametrization function, $w(x)$,  by 
\bea \label{fw}
 f_P(x)&=&\alpha'_P \log\Big(\frac{1}{w(x)}\Big),\\ \label{qw}
g_\tau(x)&=&\frac{1}{N_\tau} \frac{w'(x)}{x}  [1-w(x)]^{\tau-2}w(x)^{1 - \alpha_P(0)},
\eea
where $g_\tau(x)$ is normalized by $\int_0^1 dx \,  xg_\tau(x) = 1$ .

If we identify $x$ with the gluon LF momentum fraction, physical constraints on $w(x)$ are imposed at small and large-$x$~\cite{Ball:2016spl}.  At $x \to 0$, $w(x) \sim x$ from Regge theory~\cite{Regge:1959mz}, and  at $x \to 1$ from the inclusive-exclusive counting rule~\cite{Drell:1969km}, $g_\tau(x) \sim (1-x)^{2 \tau - 3}$, which fixes $w'(1) = 0 $. The leading $(1-x)$-exponent determined in~\cite{Sufian:2020wcv} by fitting the NNPDF gluon distribution~\cite{Ball:2017nwa} is consistent with the large-$x$ counting rule.

The gluon distribution of the proton can be expressed as the sum of contributions from all Fock states, $  g(x) = \sum_{\tau} c_\tau g_{\tau}(x)$,
where the coefficients, $c_\tau$,  represent the   normalization of each Fock component. In practice, one has to apply a truncation up to some value of $\tau$. In this study we only keep the leading term, $\tau=4$,  and  determine the coefficient $ c_{\tau=4}$ using the momentum sum rule
\begin{align}
    \int_0^1 dx\, x \big[g(x) + \sum_{q} q(x) \big] = 1,
\end{align}
where $q$ runs over all quark flavors. It also corresponds to the sum rule of the  helicity-conserving GFF $A(t)$, 
\begin{align}
A^g(0) + \sum_{q} A^q(0) = 1,
\end{align}
 which is a measure of the momentum fraction carried by each constituent. Similarly the helicity-flip GFF $B(t)$  provides a measure of the orbital angular momentum carried by each constituent of a hadron at $t = 0$. The constraint $B^g(0) + \sum_q B^q(0) = 0$ was originally derived from the equivalence principle~\cite{Teryaev:1999su} and can be formally derived Fock state by Fock state in LF quantization~\cite{Brodsky:2000ii}.

%%%%%%%%%%%%%%%%%%%%%%%%%%%%%%%
\begin{figure}[h]
    \centering
    \setlength\belowcaptionskip{-8pt}
    \includegraphics[width=1.00\linewidth]{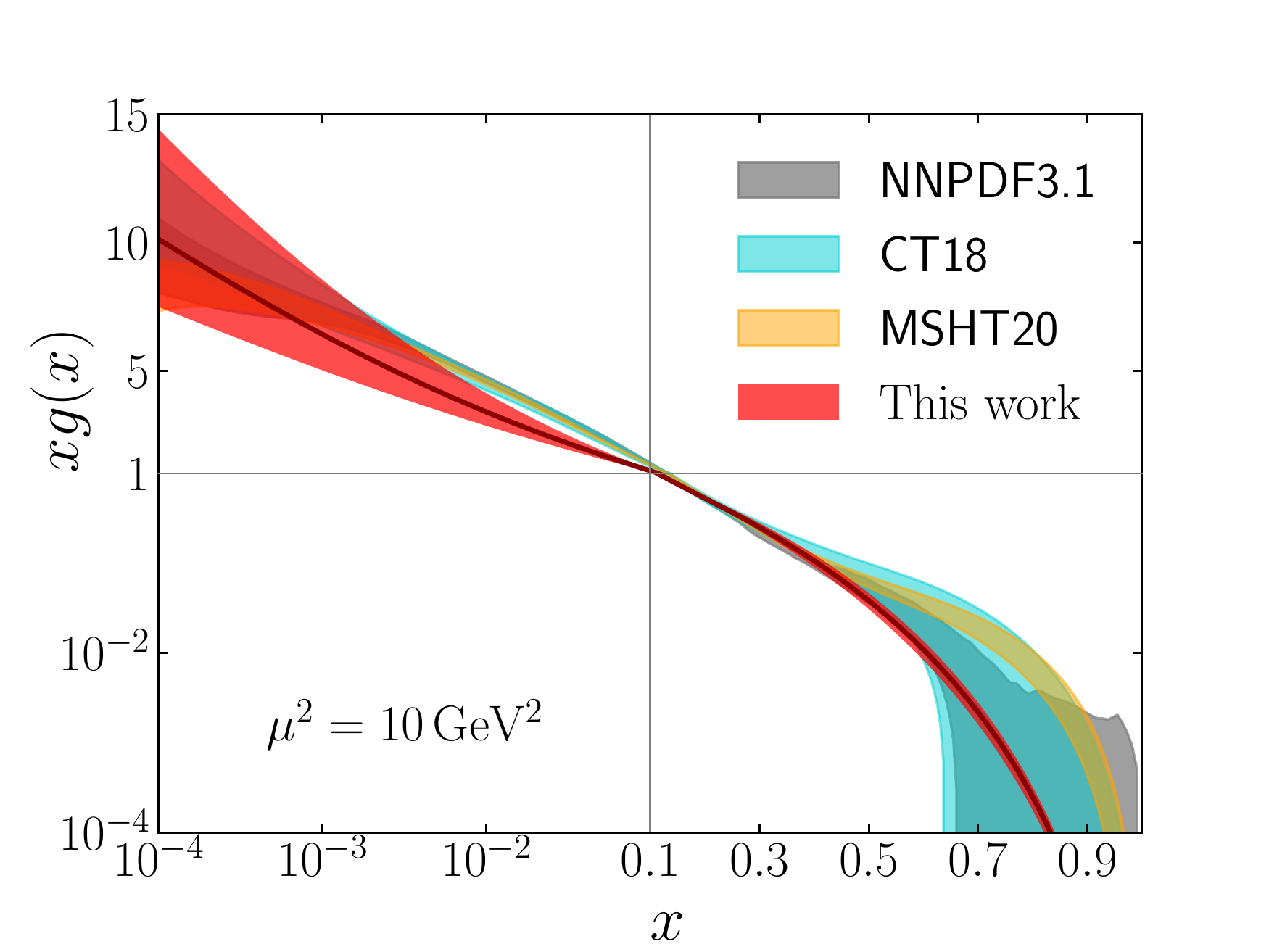}
    \includegraphics[width=1.00\linewidth]{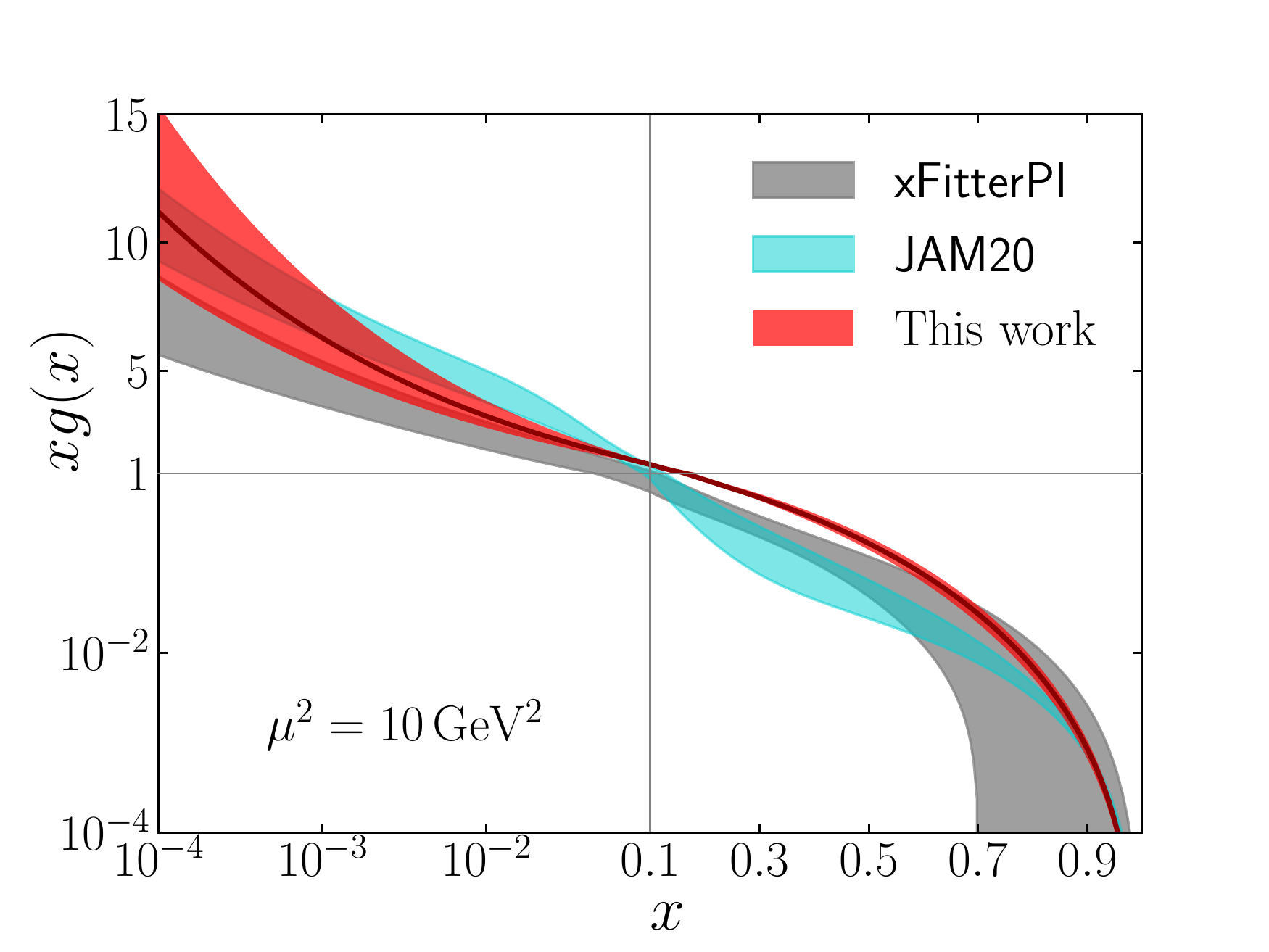}
    \caption{Comparison of the gluon distribution in the proton (up) and pion (down) with  global fits. The red bands indicate the theoretical uncertainty from the initial scale $\mu_0$ and the the gluon coefficient $c_{\tau}$ from the momentum sum rule.}
    \label{fig:PDFs}
\end{figure}
%%%%%%%%%%%%%%%%%%%%%%%%%%%%%%%

Taking the quark distributions and the reparametrization function $w(x)$ from Ref.~\cite{Liu:2019vsn}, $w(x) = x^{1-x}\exp[-b(1-x)^2]$ with $b = 0.48\pm0.04$, we determine $c_{\tau=4}=0.225  \pm  0.014$ and thus the  gluon distribution at the input scale $\mu_0 = 1.06 \pm 0.15$  GeV.  Together with the quark distributions, we evolve the gluon distribution to $\mu^2=10 \,\rm GeV^2$.  The  model results are compared in Fig.~\ref{fig:PDFs} with global analyses of  gluon  PDFs~\cite{Ball:2017nwa, Hou:2019efy, Bailey:2020ooq}. Contributions from higher Fock states are expected to be suppressed at large $x$ and may affect the overall normalization through the momentum sum rule, thus suppressing the gluon distribution at large $x$ while 
enhancing the distribution at small $x$. Our nonperturbative results falloff as $(1-x)^5$ at large $x$ from the leading
twist-four term in \req{qw}. In contrast, the perturbative QCD results incorporate a $(1-x)^4$ falloff from hard gluon transfer to the spectator quarks~\cite{Brodsky:1994kg}.

Incorporating the universality of our approach, we now  compute the gluon distribution in the pion. Similar to the case of the proton, we only consider the lowest $\tau=3$ Fock state $\ket{u\bar{d}g}$ with one constituent gluon. The  coupling of the Pomeron to the  
hadrons  depends on the vertex,  but the trajectory $ \alpha_{P}(t)$~\eqref{RP} is the same and unique to the Pomeron. 
Considering the valence quark distributions determined in Ref.~\cite{deTeramond:2018ecg},  
\begin{align}
\vert  \pi\rangle = \sqrt{0.875} \,  |u \bar{d}\rangle + \sqrt{0.125} \, |u \bar{d}q  \bar{q} \rangle,
\end{align}
we express the total quark distribution  as 
\begin{align}
q_\pi(x) &\equiv q_u(x) + q_d(x) + q_{\bar u}(x) + q_{\bar d}(x) \nn \\
&\simeq  2\times 0.875\, q_{\tau=2}(x) + 4\times 0.125\, q_{\tau=4}(x) \nn \\
& = 1.75 \, q_{\tau=2}(x) +  0.5 \, q_{\tau=4}(x).
\end{align}
Using the momentum sum rule we obtain the gluon distribution of the pion as $g_\pi (x) = (0.429\pm0.007)  \, g_{\tau=3}(x)$.  We show  in Fig.~\ref{fig:PDFs} the results evolved to $\mu^2=10\,\rm GeV^2$ and a comparison  with global analyses~\cite{Novikov:2020snp,Cao:2021aci}. 
We note that the overall normalization of the gluon distribution  from our calculation seems overestimated in comparison with some recent global analyses, which may arise from neglecting  higher Fock states  for the gluon GFF.

%%%%%%%%%%%%%%%%%%%%%%%%%%%%%%%

\section{Conclusion and outlook \lb{CaO}}

The light-front holographic extension presented in this article allows us to give a simultaneous description of the  intrinsic gluon distributions, GPD and PDF, and the gluon GFF $A_g(t)$, within a unified nonperturbative framework, for both the nucleon and pion.  The actual computations are based on the holographic coupling of the spin-two soft Pomeron to the hadron energy-momentum tensor, constrained by the Veneziano structure. The comparison of  our theoretical predictions, after DGLAP  evolution, with global analyses and recent lattice computations clearly demonstrates the predictive power of this new framework.

In this article we have only used the parameters of the soft Pomeron. For hard processes the contribution of a hard Pomeron with a much larger intercept and smaller slope also seems to be necessary~\cite{Donnachie:2002en}. However, after evolution, our results describe the full gluon distribution in accordance with phenomenological determinations, see Fig.~\ref{fig:PDFs}. This supports attempts to use the gauge/gravity duality~\cite{Maldacena:1997re, Brower:2006ea, Hatta:2007he} to describe hard and soft diffractive processes within a unified framework. The  properties of the Pomeron depend then on the kinematic regime of the scattering process~\cite{Brower:2006ea, Hatta:2007he, Dosch:2015oha}.

The gluonic and quark distributions of hadrons are shown to have significantly different effective QCD scales and sizes compared to their electromagnetic distributions. The twist-three Fock state $\vert q\bar{q} g\rangle$ in the pion ($\vert qqqg\rangle$ for the nucleon) is responsible for the intrinsic gluon distribution at the initial evolution scale where the Pomeron probe couples strongly with the constituent gluon. However, this Fock component does not contribute significantly to the quark GPD since the EM probe does not resolve the deeply bound constituent gluon~\cite{Brodsky:2011pw}, and thus it is effectively included in the twist-two $q \bar q$ state ($qqq$ for the nucleon). Our results for $A(t)$ can be extended to the other two invariant GFFs $B(t)$ and $C(t)$~\cite{Pagels:1966zza}. Of particular relevance is the coupling of the scalar Pomeron trajectory--with the  similar slope, but different intercept, to  determine the form factor $C(t)$~\cite{Mamo:2019mka, Mamo:2021krl}. It can allow us to gain further insights into the distribution of internal shear forces and pressure inside the proton and therefore of its dynamical stability.

%%%%%%%%%%%%%%%%%%%%%%%%%%%%%%%%

%%%%%%%%%%%%%%%%%%%%%%%%%%%%%%%%

\acknowledgments{

 RSS thanks Patrick Barry for providing the JAM20  gluon distribution of the pion. TL is supported in part by National Natural Science Foundation of China under Contract No. 12175117. RSS is supported  by U.S. DOE grant No. DE-FG02-04ER41302 and in part by the U.S. Department of Energy contract  No. DE-AC05-06OR23177,  under  which  Jefferson  Science  Associates, LLC, manages and operates Jefferson Lab. SJB is supported in part by the Department of Energy Contract No. DE-AC02-76SF00515.}

%%%%%%%%%%%%%%%%%%%%%%%%%%%%%%%%%%

%%%%%%%%%%%%%%%%%%%%%%%%%%%%%%%%%%

%%%%%%%%%%%%%%%%%%%%%%%%%%%%%%%%%%%%%%%%%%%%%%%%%%%%%

%%%%%%%%%%%%%%%%%%%%%%%%%%%%%%%%%%%%%%%%%%%%%%%%%%%%%

\end{document}